# Nongray EWB and WSGG Radiation Modeling in Oxy-Fuel Environments

Osama A. Marzouk[1] and E. David Huckaby[2]

[1]*U.S. Department of Energy, National Energy Technology Laboratory; and West Virginia University Research Corporation*
[2]*U.S. Department of Energy, National Energy Technology Laboratory*
*USA*

## 1. Introduction

According to a recent U.S. Greenhouse Gas Emissions Inventory (1), about 42% of 2008 $CO_2$ (a greenhouse gas) emissions in the U.S were from burning fossil fuels (especially coal) to generate electricity. The 2010 U.S. International Energy Outlook (2) predicts that the world energy generation using coal and natural gas will continue to increase steadily in the future. This results in increased concentrations of atmospheric $CO_2$, and calls for serious efforts to control its emissions from power plants through *carbon capture* technologies. Oxy-fuel combustion is a *carbon capture* technology in which the fossil fuel is burned in an atmosphere free from nitrogen, thereby reducing significantly the relative amount of $N_2$ in the flue-gas and increasing the mole fractions of $H_2O$ and $CO_2$. This low concentration of $N_2$ facilitates the capture of $CO_2$. The dramatic change in the flue composition results in changes in its thermal, chemical, and radiative properties. From the modeling point of view, existing transport, combustion, and radiation models that have parameters tuned for air-fuel combustion (where $N_2$ is the dominant gaseous species in the flue) may need revision to improve the predictions of numerical simulations of oxy-fuel combustion.

In this chapter, we consider recent efforts done to revise radiation modeling for oxy-fuel combustion, where five new radiative-property models were proposed to be used in oxy-fuel environments. All these models use the weighted-sum-of-gray-gases model (WSGGM). We apply and compare their performance in two oxy-fuel environments. Both environments consist of only $H_2O$ and $CO_2$ as mixture species, and thus there is no $N_2$ dilution, but the environments vary in the mole fractions of these two species. The first case has a $CO_2$ mole fraction of 65%, whereas the second has a $CO_2$ mole fraction of 90%. The former case is more relevant to what is referred to as *wet flue gas recycle (wet FGR)* where some flue gas is still recirculated into the furnace, but after to act as coal carrier or diluent (to temper the flame temperature). On the other hand, the second case is more relevant to what is referred to as *dry flue gas recycle (dry FGR)* where some flue gas is still recirculated into the furnace but after a stage of $H_2O$ condensation. This increases the $CO_2$ fraction in the recycled flue gas (RFG) and consequently in the final flue gas leaving the furnace and the boiler of the plant.

To highlight the influence of using an air-fuel WSGGM (a model with parameters were developed for use in air-fuel combustion) in oxy-fuel environments, the air-fuel WSGGM of Smith et al. (1982) is included as the sixth WSGGM. The WSGG solutions are accompanied by solutions using the more-rigorous exponential wide band model (EWBM) approach and the spectral line-base weighted-sum-of-gray-gases model (SLW) approach. All the solutions presented here are nongray, meaning that the radiative properties of the emitting/absorbing mixture vary across the spectrum and multiple radiative transfer equations (RTEs) are solved per spectrum. The total pressure is 1 atm (101 325 N/m$^2$).

## 2. Mathematical description

The spectral radiative transfer equation (RTE) along a path $s$ (with a unit vector $\hat{s}$) in an emitting/absorbing medium is (3; 4)

$$\frac{d\,I_\eta(s,\eta)}{d\,s} = \hat{s} \bullet \nabla I_\eta = k_\eta(s,\eta) \left( I_{b,\eta}(s,\eta) - I_\eta(s,\eta) \right) \tag{1}$$

where $\eta$ is the wavenumber (its SI unit is $1/m$), $I_\eta$ is the spectral radiative intensity (its SI unit is $\frac{W/m^2}{m\ \text{steradian}}$), $I_{b,\eta}$ is the blackbody radiative intensity, and $k_\eta$ is the spectral linear radiative absorption coefficient (its SI unit is $1/m$). From a molecular view, when $k_\eta$ is uniform along a path, $1/k_\eta$ is the mean free path traveled by a photon until it is absorbed by an electron (3; 4). From a continuum view, and from Equation (1), it can also be viewed as simply the fraction of radiation pencil absorbed over a distance of 1 meter (5). The blackbody radiative intensity, or the Planck function, ($I_{b,\eta}$) depends on the wavenumber (or wavelength), local temperature, and the refractive index of the medium. This dependence has the following form:

$$I_{b,\eta}(s,\eta) = \frac{2\,h\,c_0^2\,\eta^3}{n^2 \left( \exp\left( \frac{h\,c_0\,\eta}{k_B\,T} \right) - 1 \right)} \tag{2}$$

where $n$ is the refractive index of the medium (being unity for vacuum), $h$ is the Planck constant (in SI units, $h = 6.6261 \times 10^{-34}$ J-s), $c_0$ is the speed of light in vacuum (in SI units, $c_0 = 299\ 792\ 458$ m/s), $k_B$ is the Boltzmann constant (in SI units, $k_B = 1.3807 \times 10^{-32}$ J/K), and $T$ is the temperature. When Equation (2) is integrated over the entire spectrum, we obtain the total blackbody radiation intensity, which depends only on the medium type (through its refractive index) and the local temperature, as follows:

$$I_{b,tot}(n,T) = n^2\,\sigma\,T^4/\pi \tag{3}$$

where $\sigma$ is the Stefan-Boltzmann constant (in SI units, $\sigma = 5.67 \times 10^{-8}$ W/m$^2$-K$^4$).
In modeling, the thermal effect of radiation appears in the energy equation through a radiative source term (its SI unit is W/m$^3$), which takes the following form:

$$\text{source} = \int_{\eta=0}^{\infty} \left( \int_{4\pi} k_\eta\,I_\eta d\Omega - 4\pi\,k_\eta\,I_{b,\eta} \right) d\eta \tag{4}$$

where $\Omega$ is the solid angle (in steradian). This source term is negative when the radiation has cooling effect on the medium, as in flames and reacting flows (6; 7).
Defining the spectral direction-integrated incident radiation

$$G_\eta \equiv \int_{4\pi} I_\eta d\Omega \tag{5}$$

then Equation (4) can be re-written as

$$\text{source} = \int_{\eta=0}^{\infty} \left( k_\eta \, G_\eta - 4\pi \, k_\eta \, I_{b,\eta} \right) d\eta \tag{6}$$

In the most comprehensive approach, known as the line-by-line (LBL) approach (4), the spectrum is divided into high-resolution intervals where $k_\eta$ is approximately constant over each interval, and an RTE per direction is solved for each interval. Then, the total radiative intensity and the total radiative source term are obtained from spectral integration of the respective spectral quantities.

The spectral absorption coefficient for gaseous species is known to vary rapidly and it is far from being a smooth function of $\eta$. This is due to the fact that radiation from a hot gas (e.g., a flame) is absorbed by combustion gases only at wavenumbers at which electrons can be excited to the next discrete energy level. Therefore these gases are radiatively-transparent at certain portions of the spectrum, but become radiatively-active at other portions (8). The LBL approach for solving the radiation problem is not practical in real combustion simulations, where such approach would involve hundreds of thousands of RTEs. Alternative approaches exist where much fewer RTEs are solved to resolve the spectrum.

Of course the extreme case is to solve a single RTE per direction for the entire spectrum, assuming constant properties over the entire spectrum. This approach is referred to as *gray*. This simplifies the calculations greatly, but completely loses the spectral character of radiation through its full-spectrum averaging. In that approach, the RTE becomes

$$\frac{d \, I_{tot}}{d \, s} = \hat{s} \bullet \nabla I_{tot} = k_{gray} \left( I_{b,tot} - I_{tot} \right) \tag{7}$$

and the radiative source becomes

$$\text{source} = k_{gray} \left( G_{tot} - 4 \, \pi \, I_{b,tot} \right) \tag{8}$$

As a compromise between the formidable LBL approach and the too-coarse gray approach, we apply two other approaches where spectral variation is accounted for, but with a much lower resolution than the LBL. These approaches are the nongray WSGGM and the box model based on the EWBM. In either approach, for each direction a small number of RTEs solved, each of which covers a fraction of the spectrum where the linear absorption coefficient is considered to be constant, and where the fraction of the total blackbody radiation over that spectral portion is what acts to augment the radiation. Therefore, the RTE of the $i^{th}$ fraction is

$$\frac{d \, I_i}{d \, s} = \hat{s} \bullet \nabla I_i = k_i \left( a_i \, I_{b,tot} - I_i \right) \tag{9}$$

where the quantity $a_i$ is the fraction of the total (i.e., spectrally-integrated) blackbody radiation that belongs to the $i^{th}$ spectral fraction. The source term is

$$\text{source} = \sum_i k_i \left( G_i - a_i \, 4 \, \pi \, I_{b,tot} \right) \tag{10}$$

In the box/EWB, the spectrum partitioning is based on modeled band structure that reflects the presence of the vibration-rotation or pure-rotation bands of the emitting/absorbing species. In the nongray WSGGM, no direct partitioning of the spectrum is done, and each of the so-called fractions is a hypothetical collection of noncontiguous intervals of the spectrum having the same value of the spectral absorption coefficient. In the following subsections, we describe further the box/EWB model and the nongray WSGG model.

### 2.1 Box/EWB model

In the general box model, the erratic spectral profile of $k_\eta$ is idealized as a piecewise-constant function, with constant $k_\eta$ values over a range of $\eta$. This value can be zero over intervals of spectrum where no absorption is occurring (called the *windows*). In the present work, a piecewise-constant function of $k_\eta$ is calculated using the exponential wide band model, which idealizes each vibration-rotation band of $H_2O$ or $CO_2$ as well as the far-infrared pure-rotation band of $H_2O$ according to the block approximation (9). A block is formed between the edges of each idealized band. There are 6 vibration-rotation bands of $CO_2$, four vibration-rotation bands of $H_2O$, and a pure-rotation band of $H_2O$. The number of blocks varies depending on the width of each idealized band; which in turn depends on the fractions of $H_2O$ and $CO_2$ in the medium, its temperature, and its total pressure.

We have used a model with 22 blocks that cover the wavenumbers from $\eta$=0 to 100 000 1/cm. This corresponds to wavelengths from $\lambda$=0.1 $\mu$ m to $\infty$. Such range is wide enough to handle thermal radiation (10). Consequently, 22 RTEs per direction are solved to resolve the spectrum. This range covers more than 99.99% of the area under the Planck function at 1 500 K. The band equivalent widths are computed using the Edwards-Menard 3-regime expressions (11; 12) for the vibration-rotation bands, and using the Fleske-Tien theoretical expression (13; 15) for the pure-rotation band. The parameters for the vibration-rotation bands are those in (14) and for the pure-rotation bands are those in (15). Relating this approach to Equation (9), each $k_i$ is a block value and each $a_i$ is the fraction of the Planck function over that block. The box/EWB approach requires the specification of a mean pathlength (some characteristic length for radiation) for the problem, which is approximated as 3.6 times the volume divided by the surface areas (3). For the $12 \times 12 \times 40$ m rectangular enclosure we consider here, this value is 9.3913 m. This length was also used to obtain $k_i$ for each block from its calculated emissivity.

Figure 1 shows the idealized spectra of $k_\eta$ for the two gas compositions studied in this work (i.e., 65% $CO_2$ with 35% $H_2O$, and 90% $CO_2$ and 10% $H_2O$) at a constant temperature of 1 500 K. The corresponding blackbody emissive power ($E_{b,\eta} = \pi \, I_{b,\eta}$) is superimposed in each plot. The corresponding spectra using the wavelength $\lambda$ as the spectral variable are given in Figure 2.

Whereas both $CO_2$ and $H_2O$ are radiatively-active, as some $H_2O$ is replaced by $CO_2$ (moving from wet recycle to dry recycle), the absorption/emission of the mixture decreases (17; 18). The full listing of the linear absorption coefficients and blackbody weights of each block for both oxy-fuel environments are given in Appendix 7, Tables 9 and 10.

### 2.2 WSGGM

Despite the large reduction in the number of calculations when switching from the LBL approach to the box/EWB approach, it is still desirable to attain further reduction in the number of RTEs to be solved for the entire spectrum when performing complex combustion simulations as they involve many physical and chemical phenomena other than radiation. The WSGGM has enjoyed great popularity (4) and is utilized here as a *more-practical* approach for complex combustion modeling, whereas the aforementioned more-expensive

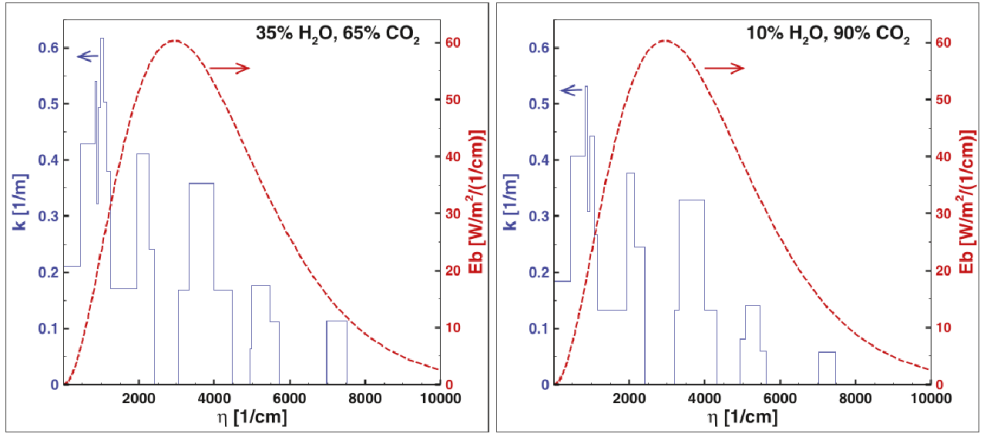

Fig. 1. Spectra (versus wavenumber) of the blackbody emissive power and the box/EWB linear absorption coefficient and at 1 500 K for two oxy-fuel environments

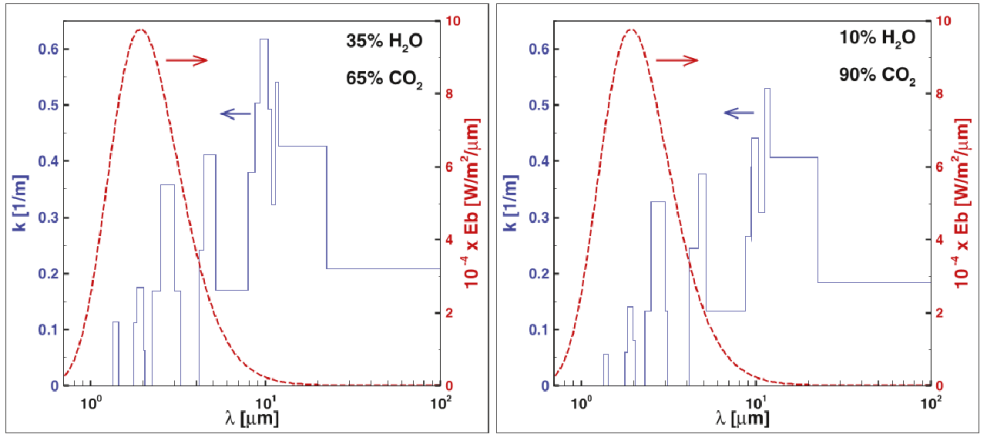

Fig. 2. Spectra (versus wavelength) of the blackbody emissive power and the box/EWB linear absorption coefficient and at 1 500 K for two oxy-fuel environments

box/EWB approach serves to provide a benchmark solution to compare with. In the nongray WSGG approach, Equation (9) is still solved as was the case in the box model, but the physical interpretation and the evaluation of the $k_i$ and $a_i$ are very different. The WSGG approach (5; 16; 19–21) is based on the presence of $N$ hypothetical gray gases; $N - 1$ are absorbing/emitting, and one is clear (no radiative emission or absorption) to represent the presence of spectral windows. Each absorbing/emitting gray gas has a constant $k_i$, and the clear gas has $k_0=0$. The fractions $a_i$ are cast as a polynomial of temperature only. The parameters of a WSGGM are the $k_i$ and the polynomial coefficients for each absorbing/emitting gray gas. There are $(N - 1) \times (M + 1)$ model parameters for $N$ gray gases and a polynomial order $M$. The parameters for a single total pressure and a single gas composition ($H_2O$ and $CO_2$ partial pressures) are calculated through an optimization process.

The optimization requires a set of emissivities for a range of temperatures and pathlengths at these total pressure and gas composition.

When used to calculate the total emissivity (either during the model coefficient optimization process or for evaluating the total emissivity with fixed model coefficients), the WSGGM returns a weighted sum of individual emissivities of the hypothetical absorbing/emitting gray gases, i.e.

$$\epsilon_{tot} = \sum_{i=1}^{N-1} a_i(T) \left(1 - \exp\left[-K_{p,i} \, PL\right]\right) \tag{11a}$$

$$\text{where} \quad a_i(T) = \sum_{j=1}^{M+1} b_{ij} \, \left(T/\hat{T}\right)^{j-1} \tag{11b}$$

where $\epsilon_{tot}$ is the total emissivity (dimensionless), $PL$ is pressure-pathlength, $L$ is the mean pathlength, $K_{p,i}$ are the pressure absorption coefficients for the $N-1$ absorbing/emitting gray gases, $a_i$ are the blackbody weights for these absorbing/emitting gray gases, $b_{ij}$ are the coefficients for a polynomial of degree $M$ in $T/\hat{T}$, and $\hat{T}$ is a scaling temperature that aids in the minimization process.

When the WSGGM is used to perform nongray calculations for use in Equation (9), the weights $a_i$ are also evaluated from the temperature polynomial in Equation (11b); the $i^{th}$ linear absorption coefficient is evaluated as

$$k_i = K_{p,i} \, P \tag{12}$$

where $P$ is the sum of the partial pressures of $H_2O$ and $CO_2$ (in units consistent with those of $K_{p,i}$). A total of $N$ RTEs are solved per direction to resolve the spectrum. In the WSGG models considered here, $N$ takes the value of 4 or 5, which is a considerable reduction in computations compared to the box/EWB procedure described in subsection 2.1.

Table 1 compares the characteristics of the WSGG models which we consider. The first five WSGG models have been optimized for oxy-fuel combustion, whereas the last was developed for air-fuel combustion. Its inclusion in the study is a method to estimate the errors in radiation modeling when applying air-fuel WSGG models in oxy-fuel combustion simulations.

All models shown in Table 1 have mode parameters at finite sets of gas compositions, except for the 2011 model of Johansson et al. (24) where the model parameters are expressed as continuous functions of the molar ratio $H_2O/CO_2$. We perform piecewise-linear interpolation/extrapolation using the molar ratio $H_2O/(H_2O+CO_2)$ as an independent variable to apply the model at arbitrary gas compositions (18; 26). Marzouk and Huckaby (18) compared this technique to the piecewise-constant technique and recommended the former based on gray radiation modeling of non-isothermal media. The full listing of the linear absorption coefficients and blackbody weights for the gray gases of the 6 WSGG models for both oxy-fuel environments is given in Appendix 7, Tables 11-16. Notice that for either oxy-fuel environment, the clear-gas weight $(a_0)$ in the air-fuel WSGGM (28) is higher than its counterpart in all the oxy-fuel WSGG models. This acts to reduce the radiative participation of the gaseous mixture.

| Ref. | $N$ | T-poly | Num. sets | $\hat{T}(K)$ | T range (K) | PL range | Training data |
|------|-----|--------|-----------|--------------|-------------|----------|---------------|
| (22) | 4 | quadratic | 2 | 1 200 | $500-2\,500$ | $0.01-60$ bar-m | SNBM (23) |
| (22) | 5 | quadratic | 2 | 1 200 | $500-2\,500$ | $0.01-60$ bar-m | SNBM (23) |
| (24) | 5 | quadratic | N/A | 1 200 | $500-2\,500$ | $0.01-60$ bar-m | SNBM (23) |
| (25) | 4 | linear | 3 | 1 | $1\,000-2\,000$ | $0.005-10$ atm-m | empirical correlation (26) |
| (27) | 5 | cubic | 7 | 1 200 | $500-3\,000$ | $0.001-60$ atm-m | EWBM (9; 14; 15) |
| (28) | 4 | cubic | 5 | 1 | $300-3\,000$ | $0.001-10$atm-m | EWBM (9; 14; 15) |

[a] at compositions: $H_2O:CO_2$ = 11.1%:88.9% and 50%:50% by mole

[b] *SNBM* is statistical narrow band model

[c] $K_{p,i}$ are expressed as linear functions of the $H_2O/CO_2$ molar ratio, and $b_{ij}$ as quadratic functions of it

[d] at compositions: $H_2O:CO_2$ = 10%:10% (80% $N_2$), 33%:66%, and 10%:90% by mole

[e] implied from the empirical correlation used for the training data

[f] at compositions: $H_2O:CO_2$ = 11.1%:88.9%, 20%:80%, 33.3%:66.7%, 42.9%:57.1%, 50%:50%, 66.7%:33.3, 80%:20 by mole (3 others sets are given but not for oxy-fuel environments)

[g] at compositions: $H_2O:CO_2$ = $\rightarrow$ 0:0 (diluent is $N_2$), 10%:10% (80% $N_2$), 20%:10% (70% $N_2$), 0:$\rightarrow$ 0, 0%:100% (diluent is $N_2$) by mole.

Table 1. Summary of the 6 WSGG models considered here (5 oxy-fuel and 1 air-fuel)

## 3. Test cases

In coupled combustion simulations, different sub-models interact and thus it becomes difficult to examine the independent response of a particular sub-model. It is advantageous to isolate the radiation modeling when examining different solution approaches, which is what we have followed here. The two test problems to be presented in this section correspond to a stagnant homogeneous isothermal gas mixture. Only the radiative intensity is allowed to vary, thereby eliminating cross-model interactions which could make it difficult to judge the performance of the performance of the particular radiation model from the simulation results. Since our primary goal is to study the performance of the different oxy-fuel WSGG models when used in oxy-fuel environments, we considered two idealized oxy-fuel product gas compositions. Both environments have an atmospheric total pressure, which is also the sum of the partial pressures of $H_2O$ and $CO_2$ (thus, no $N_2$ dilution, which is relevant to oxy-fuel operations). The only difference between the two environments is the gaseous composition, which is summarized in Table 2. In both environments, the $CO_2$ mole fraction is higher than the fraction of the $H_2O$. However, the second environment features dominance of $CO_2$ (9 times $H_2O$), which is more relevant to dry-recycle oxy-fuel operations.

| Test case | $H_2O$ mole fraction | $CO_2$ mole fraction | Total pressure (atm) | Temperature (K) |
|-----------|----------------------|----------------------|----------------------|-----------------|
| case 1 | 35% | 65% | 1 | 1 500 |
| case 2 | 10% | 90% | 1 | 1 500 |

Table 2. Summary of the 2 studied oxy-fuel environments

The geometry of both problems is a large rectangular enclosure, with dimensions $12 \times 12 \times 40$ m. The medium temperature is 1 500 K. The temperature of the walls is kept at 750 K, with an emissivity of 0.725. This configuration was proposed by Krishnamoorthy et al. (25) to roughly represent the dimensions of a full-scale 300 MW front-wall-fired, pulverized-coal, utility boiler (29). The domain is discretized with a uniform mesh of $27 \times 27 \times 82$ cells, resulting in a total of 59 778 hexahedral cells.

## 4. Results

### 4.1 Numerical settings

The box/EWB model and each of the 6 WSGG models are applied to each of the 2 oxy-fuel environments. As mentioned in subsections 2.1 and 2.2, there are 22 RTEs per direction to resolve the spectrum for the box/EWB approach, and either 4 or 5 RTEs per direction to resolve the spectrum for the WSGG approach. We use the finite-volume method for the both the spatial and directional discretizations. As mentioned earlier in section 3, the enclosure is discretized into 59 778 cells. In each cell, the 3D angular space of $4\pi$ is divided into 128 angular divisions. A coarse representation of such angular discretization for a hemisphere (angular space of $2\pi$) is shown in Figure 3. We have performed sensitivity analyses to check the suitability of both linear and angular resolutions by comparing a solution obtained using the aforementioned ones with a solution obtained using a finer linear resolution ($33 \times 33 \times 110 = 119$ 790 cells) while keeping the angular resolution unchanged; and with a solutions obtained using a finer angular resolution (200 divisions) while keeping the linear resolution unchanged. In both situations, the solutions are nearly identical, and thus the adopted resolutions are considered sufficient. The nongray radiation simulation is performed iteratively using the computational fluid dynamics software ANSYS FLUENT 13.0 (30). None of the radiative-property models described here are available in the standard release. We implemented each method through a user-defined function that is complied and linked to the software for run-time access.

### 4.2 Order of presentation

In the four subsequent subsections, the solutions of the radiative solution for the two oxy-fuel environments are presented. We first start in subsection 4.3 with 2D flooded contours of the radiative source term (in kW/m$^3$) along the 12×40 vertical midplane (the symmetry plane midway between the two vertical side walls separated by a distance of 12 m). Due to the symmetry of the problem, this plane should be identical to the horizontal symmetry plane. Next, the 1D profiles of this radiative source term along the centerline of the enclosure (i.e., the 40-m longitudinal line passing through the geometric center of the 12×12 cross-section of the enclosure) are presented in subsection 4.4. In these profiles, we also include published results (25) using the SLW approach.

The SLW approach (originally proposed by Denison and Webb (31)) is a more-rigorous implementation of the WSGGM. The individual gray gases now have a physical meaning and direct mathematical relationship with the absorption spectrum (in terms of the absorption cross-section, whose SI unit is m$^2$/mol). The range of the absorption coefficient is divided into segments, each of which represents an absorbing/emitting gray gas. In addition, there is one clear gas (as in the WSGG approach). The segmentation of the range of absorption coefficient is typically done such that their logarithmic values are equally spaced. For each segment (i.e., each absorbing/emitting gray gas), a logarithmic average absorption cross-section $C_i$ is assigned, and the corresponding blackbody weight $a_i$ is evaluated to be the fraction of the Planck function that belongs to the range of absorption coefficient of the segment represented by the $i^{th}$ gray gas. The linear absorption coefficient for a species is related to the absorption cross-section by the species molar concentration (its SI units is kmol/m$^3$). The exact implementation of this method would require the processing of a high-resolution spectrum (which incurs the processing of millions of spectral data points at high combustion or flue temperatures), the computations are highly simplified by utilizing a fitted hyperbolic tangent function for the cumulative distribution of the absorption cross-section, which is known as the absorption-line blackbody distribution function (ALBDF) (32; 33).

Several different approaches have been developed to apply the SLW method to multicomponent gas mixtures. The approaches are derived using different assumptions and vary in computational cost and accuracy. For the SLW solution we include here, the absorption cross-section domains of $H_2O$ and $CO_2$ were individually discretized into 20 logarithmically-spaced intervals between $3 \times 10^{-5}$ m$^2$/mol and 120 m$^2$/mol for $H_2O$, and between $3 \times 10^{-5}$ m$^2$/mol and 600 m$^2$/mol for $CO_2$. The analytical expressions for the absorption-line blackbody distribution functions of $H_2O$ (32) and $CO_2$ (33) were used to compute the blackbody weights of each gray gas. The multiplication method (34) was used to handle the presence of a mixture. Implied in this method, is the assumption that the absorption cross-sections of $H_2O$ and $CO_2$ are statistically independent. The number of RTEs per direction was 21 (one RTE per each of the 20 gray gases plus an RTE for the clear gas). The SLW calculations were performed using the T4 angular quadrature (35), and using the same spatial resolution we employed for the other two approaches (namely 27×27×82) and using a similar angular resolution (128 directions).

Unlike the box/EWB and WSGG solutions, in which we use the angular finite-volume method for treating the angular dependence of radiation, the SLW solutions were obtained using the discrete-ordinate method. Whereas both methods have some similarity, the angular finite-volume method conserves the radiative energy (4) and thus is considered a more accurate method for handling the directional dependence of radiation. In addition, the analytical fits for ALBDF of $H_2O$ and $CO_2$ are based on an extension of an old version (1991/1992) of the spectral database HITRAN (36). This database was assembled for a (low) temperature of 296 K and thus when applied at high temperatures the absorption of the medium will be underpredicted because many *hot lines* (i.e., transitions from excited vibration levels) are missing (4; 32; 37). Whereas a procedure (37) was followed to extend the original database by generating *hot-line* estimates from *cold-lines* (i.e., transitions from the ground level), Modest (4) showed that these analytical expressions result in nontrivial deviations from LBL calculations at 2 000 K. However, at 1 000 K, they are in good agreement with the LBL solution. On the other hand, the SLW approach does not require the specification of a pathlength as the EWB approach.

In the legends, the different WSGG models are designated by the total number of radiating and clear gases (either 4 or 5) and the temperature-polynomial order (linear, quadratic, or cubic). Since the WSGGM in reference (22) and the one in reference (24) have the same number of gray gases and the same polynomial order (5 gases and quadratic polynomials), we add a suffix *(cont)* to the WSGGM in reference (24) to highlight that its parameters are continuous functions of the $H_2O/CO_2$ molar ratio. Also, the air-fuel WSGGM here (28) is further distinguished by adding the suffix *(air)* to its legend entry.

The following subsection, number 4.5, shows also 1D profiles, but for the to-wall radiative flux (in $kW/m^2$) along the 40-m longitudinal midline of the top $12 \times 40$ wall of the enclosure. Due to the symmetry of the problem, this should be identical to any midline on the other three $12 \times 40$ walls.

The final subsection, number 4.6, is dedicated to the area-integrated radiative heat transfer rate to the walls (in MW). This subsection provides a quantitative measure of the variation among the different solutions with regard to the total radiative heat transfer rate (in MW) to the walls of the enclosure. The area-integrated heat transfer is an important quantity when we are concerned about the operation of the furnace unit within the boiler, as this effects steam generation rate. The average radiative heat flux is calculated from this quantity by dividing area-integrated heat transfer rate by the total surface area of the walls ($2\ 208\ m^2$), and is also included in the comparison tables. This quantity provides a geometry-independent measure of the radiative heat load in oxy-fuel furnaces. The deviations from the benchmark box/EWB solution are also included. One table is provided per oxy-fuel environment.

### 4.3 Radiative-source contours

Slices of the radiative source term along the $12 \times 40$ plane of symmetry are shown in Figure 4. Each figure corresponds to a different model, with a plot for each of the two oxy-fuel environments. The number and values of the contour levels are the same for the plots. The double-symmetric pattern in all plots is expected. The calculated negative value of the radiative heat source would *drive* the temperature field to lower values in a coupled simulation. The radiative source is smallest near the colder-than-medium walls; it increases steeply and becomes nearly flat over a large portion of the plane. Notice that this value is very similar for both environments. The box/EWB solution exhibits a smaller decrease of the radiative source near the walls than the WSGG solutions. For both environments, the air-fuel WSGGM (28) and the 5-gas/cubic WSGGM (27) show noticeable overprediction of the radiative source, which indicates a weaker influence of radiation on the thermal field.

### 4.4 Radiative-source profiles

Profiles of the radiative source term along the longitudinal centerline of the 3D enclosure are compared in Figure 5 for the two oxy-fuel environments. As mentioned in subsection 4.2, we also add published profiles (25) predicted using the SLW approach. For both environments, the flat portion of the radiative-source curve is smallest in the case of the box/EWB solution. As suggested from the 2D contours in the previous subsection, the air-fuel and the 5-gas/cubic WSGGM solutions show noticeable overprediction of the radiative source, with the air-fuel solution being the worst.

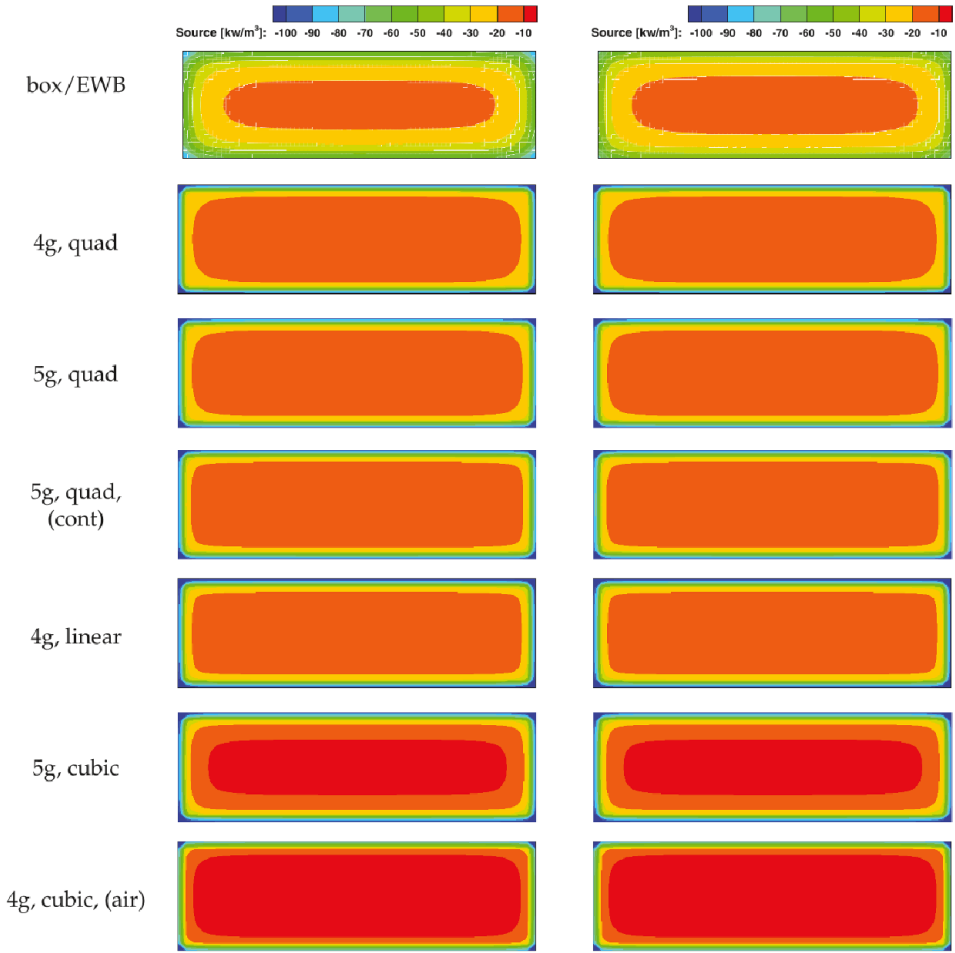

Fig. 4. Midplane radiative source (left: 65%$CO_2$; right: 90%$CO_2$)

Tables 5 and 6 list the values of the radiative source at the middle of the profiles (which corresponds to the centroid of the 3D enclosure) for the various solutions, with the relative deviation from the box/EWB solution, computed as

$$\text{Percent error} = \frac{\text{SLW/WSGG} - \text{box/EWB}}{\text{box/EWB}} \times 100\% \tag{13}$$

For the SLW and oxy-fuel WSGGM, the errors have decreased for the high-$CO_2$-fraction case, whereas this error increased in the case of the air-fuel solution. For the air-fuel solution, the errors are very large, being around 80%. Further, if the solutions are ranked by error, we get the same ordering for both oxy-fuel environments.

| Solution method | Radiative source at the centroid (kW/m$^3$) | %Error (relative to box/EWB) |
|---|---|---|
| box/EWB | -15.91 | 0.00% |
| SLW | -13.24 | +16.80 % |
| 4g, quadratic | -14.67 | + 7.80% |
| 5g, quadratic | -10.70 | +32.73% |
| 5g, quadratic, (cont) | -10.96 | +31.09% |
| 4g, linear | -11.95 | +24.88% |
| 5g, cubic | - 7.53 | +52.66% |
| 4g, cubic, (air) | - 3.22 | +79.73% |

Table 3. Radiative source term at the centroid for the oxy-fuel environment with 65%$CO_2$

| Solution method | Radiative source at the centroid (kW/m$^3$) | %Error (relative to box/EWB) |
|---|---|---|
| box/EWB | -15.15 | 0.00% |
| SLW | -13.53 | +10.75% |
| 4g, quadratic | -14.64 | + 3.37% |
| 5g, quadratic | -11.05 | +27.09% |
| 5g, quadratic, (cont) | -11.47 | +24.32% |
| 4g, linear | -11.61 | +23.37% |
| 5g, cubic | - 7.62 | +49.71% |
| 4g, cubic, (air) | - 2.52 | +83.40% |

Table 4. Radiative source term at the centroid for the oxy-fuel environment with 90%$CO_2$

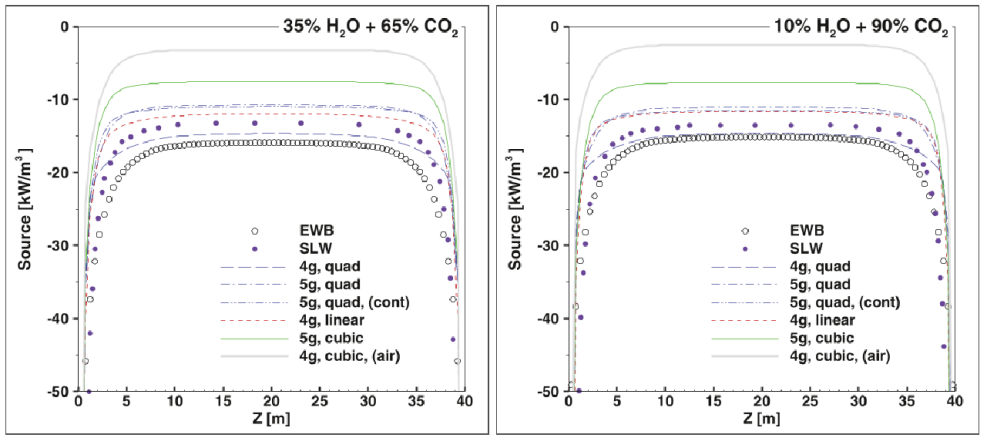

Fig. 5. Centerline radiative source for 2 oxy-fuel environments

**4.5 Radiative-flux profiles**

The profiles of the radiative flux along the symmetry line of the $12 \times 40$ top wall for the two oxy-fuel environments are shown in Figure 6. We notice that the wall radiative flux is significantly more sensitive to the change in mixture composition than the centerline radiative source (see Figure 5). When the $CO_2$ content increased, the radiative flux decreased. This is consistent with the decrease in total emissivity (18) and the changes in the idealized spectra of the linear absorption coefficient shown in Figure 1. We also notice that the relative deviations between the various WSGG models and the box/EWB predictions for the radiative flux differ from the deviations reported for the centerline radiative source. In particular, the 5-gas/cubic WSGGM (27) that showed noticeable error in the radiative source, has excellent agreement (-0.70%) with the box/EWB solution in the wet-recycle oxy-fuel environment and good agreement (-2.67%) in the dry-recycle environment. The radiative flux at the center point ($Z$=20 m) of the profiles in Figure 6 and their relative errors with respect to the box/EWB are compared in Tables 5 and 6 for the wet-recycle and dry-recycle oxy-fuel environments, respectively. All the WSGGM solutions are within 6.1% error (some underpredict and others overpredict) for both oxy-fuel environments, whereas the air-fuel WSGGM exhibits underprediction of 19.9% for the wet-recycle environment. For the dry-recycle environment, this underprediction jumps to 33.9%.

| Solution method | Wall-center's radiative flux $(kW/m^2)$ | %Error (relative to box/EWB) |
|---|---|---|
| box/EWB | 113.98 | 0.00% |
| 4g, quadratic | 119.94 | + 5.23% |
| 5g, quadratic | 119.96 | + 5.25% |
| 5g, quadratic, (cont) | 113.85 | −0.12% |
| 4g, linear | 116.33 | + 2.06% |
| 5g, cubic | 113.19 | − 0.70% |
| 4g, cubic, (air) | 91.32 | −19.88% |

Table 5. Radiative flux at top-wall center for the oxy-fuel environment with 65%$CO_2$

| Solution method | Wall-center's radiative flux $(kW/m^2)$ | %Error (relative to box/EWB) |
|---|---|---|
| box/EWB | 97.22 | 0.00% |
| 4g, quadratic | 99.67 | + 2.52% |
| 5g, quadratic | 95.83 | −1.43% |
| 5g, quadratic, (cont) | 94.37 | −2.93% |
| 4g, linear | 103.11 | + 6.05% |
| 5g, cubic | 94.63 | −2.67% |
| 4g, cubic, (air) | 64.30 | −33.87% |

Table 6. Radiative flux at top-wall center for the oxy-fuel environment with 90%$CO_2$

**4.6 Wall radiative heat transfer**

The area-integrated wall radiative heat flux results are compared for all the solutions in Table 7 for the wet-recycle environment and in Table 8 for the dry-recycle environment. Consistent with the profiles in the preceding subsection, the air-fuel WSGGM underpredicts the heat transfer for both environments. Although the relative error with respect to the box/EWB is smaller than the error recorded for the 1D flux profile, the relative error for the dry-recycle environment is larger than the relative error for the wet-recycle environment. All the oxy-fuel WSGG models overpredict the heat transfer, but the error is within 10.4%.

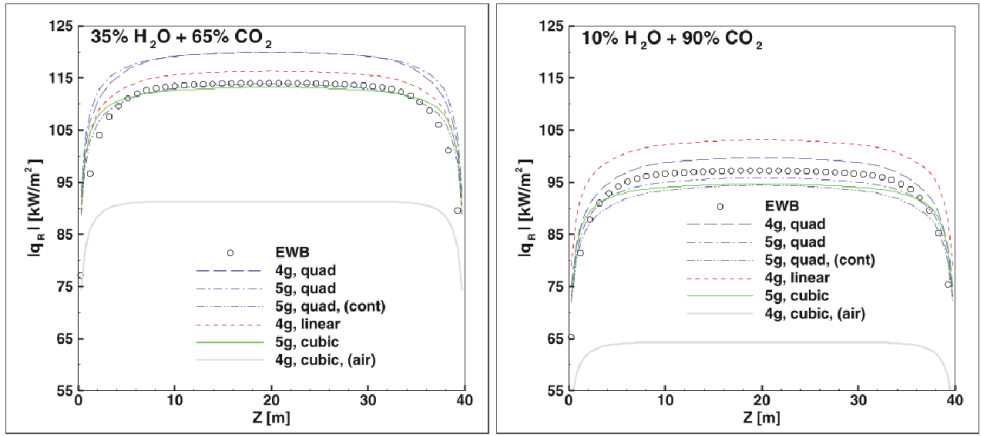

Fig. 6. Radiative flux along the midline of the $12 \times 40$ top wall

| Solution method | Wall radiative heat transfer (MW) | Average wall radiative flux (kW/m$^2$) | %Error (relative to box/EWB) |
|---|---|---|---|
| box/EWB | 224.74 | 101.78 | 0.00% |
| 4g, quadratic | 244.18 | 110.59 | + 8.65% |
| 5g, quadratic | 246.48 | 111.63 | + 9.67% |
| 5g, quadratic, (cont) | 233.99 | 105.97 | + 4.12% |
| 4g, linear | 238.15 | 107.86 | + 5.97% |
| 5g, cubic | 235.43 | 106.62 | + 4.76% |
| 4g, cubic, (air) | 191.63 | 86.79 | −14.73% |

Table 7. Wall radiative heat transfer for the oxy-fuel environment with 65%$CO_2$

## 5. Conclusions

We performed nongray radiation calculations of two radiation problems in homogeneous isothermal media.     The first medium is typical of wet-recycle oxy-fuel combustion environment, with a molar composition of 65% $CO_2$ and 35% $H_2O$; whereas the second approximates a dry-recycle environment, with a molar composition of 90% $CO_2$ and 10% $H_2O$. The domain was a $12 \times 12 \times 40$ m rectangular enclosure at 1 500 K. For each environment, we generated reference solutions using the box model based on the exponential wide band approach.     We also calculated solutions using five (recent) oxy-fuel and one (older) air-fuel weighted-sum-of-gray-gases models that were proposed in the literature. Comparing different qualitative and quantitative radiative characteristics from the obtained solutions, we see that significant improvements in predictive capability can be obtained using an oxy-WSGGM. Using the air-fuel model would result in appreciable underprediction of the local and area-integrated radiative heat flux to the wall, and in an overprediction of temperatures due to the underprediction of the heat loss due to radiation. The errors become more pronounced for the high-$CO_2$-concentration case, which is relevant to dry-recycle oxy-fuel combustion. The radiative heat flux was much more sensitive to the gas composition than the radiative source term. For the oxy-fuel WSGG models, no particular model was clearly superior. This suggests that the model used for a particular combustion problem should be selected based on the simplicity of the model and the consistency between the operating regime of the target system and the regime of the training data.

| Solution method | Wall radiative heat transfer (MW) | Average radiative flux (kW/m$^2$) | wall %Error (relative to box/EWB) |
|---|---|---|---|
| box/EWB | 190.54 | 86.30 | 0.00% |
| 4g, quadratic | 200.62 | 90.86 | + 5.29% |
| 5g, quadratic | 194.48 | 88.08 | + 2.06% |
| 5g, quadratic, (cont) | 191.72 | 86.83 | + 0.62% |
| 4g, linear | 210.34 | 95.26 | +10.39% |
| 5g, cubic | 194.76 | 88.21 | + 2.21% |
| 4g, cubic, (air) | 134.70 | 61.00 | −29.31% |

Table 8. Wall radiative heat transfer for the oxy-fuel environment with 90%$CO_2$

## 6. Acknowledgments

This technical effort was performed in support of the National Energy Technology Laboratory's ongoing research in $CO_2$ Capture in the Existing Plants Emissions and Capture (EPEC) Technology Program. Dr. Marzouk activities were funded under the RES contract DE-FE0004000. The authors appreciate the help of Dr. Chungen Yin (Aalborg University, Denmark) in implementing the EWBM.

## 7. Appendix

### A. Idealized spectra for the box/EWB approach

This appendix presents numerically the idealized spectra of the linear absorption coefficients $k_i$ and the corresponding blackbody weights $a_i$ that were computed from the EWB approach for each of the two oxy-fuel environments. The values are used when solving the RTEs given in Equation (9).

### B. WSGG linear absorption coefficients and blackbody weights

Analogous to the tabulation in Appendix 7, the computed linear absorption coefficients and the corresponding weights for the gray gases are given in this appendix for all the 6 WSGG models for each of the two oxy-fuel environments. These values are used when solving the

| $i$ | $\eta$ (1/cm) | $k_i(1/m)$ | $a_i$ | $i$ | $\eta$ (1/cm) | $k_i(1/m)$ | $a_i$ |
|---|---|---|---|---|---|---|---|
| 1 | $0.00 - 448.47$ | 0.2097897 | 0.00346421 | 12 | $2\,410.00 - 3\,048.95$ | 0.0000000 | 0.13252660 |
| 2 | $448.47 - 845.47$ | 0.4276537 | 0.01647795 | 13 | $3\,048.95 - 3\,334.04$ | 0.1683664 | 0.05929771 |
| 3 | $845.47 - 885.53$ | 0.5403818 | 0.00261648 | 14 | $3\,334.04 - 3\,985.96$ | 0.3584664 | 0.12793486 |
| 4 | $885.53 - 921.00$ | 0.3225178 | 0.00247182 | 15 | $3\,985.96 - 4\,471.05$ | 0.1683664 | 0.08419003 |
| 5 | $921.00 - 969.29$ | 0.4931395 | 0.00360145 | 16 | $4\,471.05 - 4\,929.89$ | 0.0000000 | 0.06916402 |
| 6 | $969.29 - 1\,074.53$ | 0.6162794 | 0.00879137 | 17 | $4\,929.89 - 4\,982.64$ | 0.0633557 | 0.00727657 |
| 7 | $1\,074.53 - 1\,150.71$ | 0.5035513 | 0.00716945 | 18 | $4\,982.64 - 5\,470.11$ | 0.1754496 | 0.06076172 |
| 8 | $1\,150.71 - 1\,258.43$ | 0.3804114 | 0.01127985 | 19 | $5\,470.11 - 5\,717.36$ | 0.1120939 | 0.02650485 |
| 9 | $1\,258.43 - 1\,944.35$ | 0.1706217 | 0.09986805 | 20 | $5\,717.36 - 6\,975.54$ | 0.0000000 | 0.09647500 |
| 10 | $1\,944.35 - 2\,279.00$ | 0.4116662 | 0.06230296 | 21 | $6\,975.54 - 7\,524.46$ | 0.1135142 | 0.02613845 |
| 11 | $2\,279.00 - 2\,410.00$ | 0.2410445 | 0.02595104 | 22 | $7\,524.46 - 100\,000$ | 0.0000000 | 0.06573557 |

Table 9. Idealized box/EWB spectrum for the oxy-fuel environment with 65% $CO_2$

| $i$ | $\eta$ (1/cm) | $k_i(1/m)$ | $a_i$ | $i$ | $\eta$ (1/cm) | $k_i(1/m)$ | $a_i$ |
|---|---|---|---|---|---|---|---|
| 1 | $0.00 - 440.64$ | 0.1841169 | 0.00329548 | 12 | $2\,410.00 - 3\,193.84$ | 0.0000000 | 0.16280362 |
| 2 | $440.64 - 839.06$ | 0.4063678 | 0.01624495 | 13 | $3\,193.84 - 3\,319.79$ | 0.1324004 | 0.02608835 |
| 3 | $839.06 - 893.36$ | 0.5304856 | 0.00355165 | 14 | $3\,319.79 - 4\,000.21$ | 0.3281601 | 0.13349028 |
| 4 | $893.36 - 964.10$ | 0.3082347 | 0.00513915 | 15 | $4\,000.21 - 4\,326.16$ | 0.1324004 | 0.05756875 |
| 5 | $964.10 - 1\,059.44$ | 0.4417534 | 0.00785136 | 16 | $4\,326.16 - 4\,931.64$ | 0.0000000 | 0.09340562 |
| 6 | $1\,059.44 - 1\,067.07$ | 0.2576366 | 0.00067467 | 17 | $4\,931.64 - 5\,073.92$ | 0.0809341 | 0.01929669 |
| 7 | $1\,067.07 - 1\,080.94$ | 0.3902864 | 0.00124357 | 18 | $5\,073.92 - 5\,468.36$ | 0.1404420 | 0.04830074 |
| 8 | $1\,080.94 - 1\,155.90$ | 0.2661685 | 0.00710566 | 19 | $5\,468.36 - 5\,626.08$ | 0.0595080 | 0.01724289 |
| 9 | $1\,155.90 - 1\,930.42$ | 0.1326498 | 0.10819446 | 20 | $5\,626.08 - 7\,033.12$ | 0.0000000 | 0.10905531 |
| 10 | $1\,930.42 - 2\,132.93$ | 0.3772371 | 0.03674431 | 21 | $7\,033.12 - 7\,466.88$ | 0.0571616 | 0.02063116 |
| 11 | $2\,132.93 - 2\,410.00$ | 0.2445874 | 0.05394936 | 22 | $7\,466.88 - 100\,000$ | 0.0000000 | 0.06812197 |

Table 10. Idealized box/EWB spectrum for the oxy-fuel environment with 90% $CO_2$

RTEs given in Equation (9). The linear absorption coefficient for the clear gas is $k_0=0$; its blackbody weight ($a_0$) is obtained from the requirement that $a_0 = 1 - \sum_{i=1}^{N-1} a_i$.

| | 65% $CO_2$ | | 90% $CO_2$ | |
|---|---|---|---|---|
| $i$ | $k_i(1/m)$ | $a_i$ | $k_i(1/m)$ | $a_i$ |
| 0 | 0 | 0.29433 | 0 | 0.37459 |
| 1 | 0.11695 | 0.41272 | 0.09837 | 0.41704 |
| 2 | 2.51559 | 0.23307 | 2.66557 | 0.15639 |
| 3 | 70.56945 | 0.05988 | 88.92354 | 0.05198 |

Table 11. Linear absorption coefficients and blackbody weights for the 4-gas/quadratic WSGGM in (22) – Two oxy-fuel environments

| | 65% $CO_2$ | | 90% $CO_2$ | |
|---|---|---|---|---|
| $i$ | $k_i(1/m)$ | $a_i$ | $k_i(1/m)$ | $a_i$ |
| 0 | 0 | 0.26177 | 0 | 0.31687 |
| 1 | 0.05677 | 0.30533 | 0.04006 | 0.33408 |
| 2 | 0.58148 | 0.25560 | 0.41427 | 0.20004 |
| 3 | 5.64642 | 0.13281 | 5.18028 | 0.10602 |
| 4 | 100.07946 | 0.04449 | 123.52189 | 0.04298 |

Table 12. Linear absorption coefficients and blackbody weights for the 5-gas/quadratic WSGGM in (22) – Two oxy-fuel environments

| | 65% $CO_2$ | | 90% $CO_2$ | |
|---|---|---|---|---|
| $i$ | $k_i(1/m)$ | $a_i$ | $k_i(1/m)$ | $a_i$ |
| 0 | 0 | 0.28678 | 0 | 0.34849 |
| 1 | 0.06146 | 0.33543 | 0.05633 | 0.36697 |
| 2 | 0.86869 | 0.23910 | 0.87767 | 0.17687 |
| 3 | 9.13846 | 0.10048 | 9.82222 | 0.07032 |
| 4 | 116.15385 | 0.03822 | 131.11111 | 0.03734 |

Table 13. Linear absorption coefficients and blackbody weights for the 5-gas/quadratic WSGGM in (24) – Two oxy-fuel environments

| | 65% $CO_2$ | | 90% $CO_2$ | |
|---|---|---|---|---|
| $i$ | $k_i(1/m)$ | $a_i$ | $k_i(1/m)$ | $a_i$ |
| 0 | 0 | 0.31064 | 0 | 0.32763 |
| 1 | 0.09370 | 0.33218 | 0.06288 | 0.35561 |
| 2 | 1.08144 | 0.25582 | 1.02333 | 0.22449 |
| 3 | 99.99991 | 0.10136 | 100.00000 | 0.09227 |

Table 14. Linear absorption coefficients and blackbody weights for the 4-gas/linear WSGGM in (25) – Two oxy-fuel environments

| | 65% $CO_2$ | | 90% $CO_2$ | |
|---|---|---|---|---|
| $i$ | $k_i(1/m)$ | $a_i$ | $k_i(1/m)$ | $a_i$ |
| 0 | 0 | 0.31812 | 0 | 0.39788 |
| 1 | 0.05225 | 0.22831 | 0.05105 | 0.23703 |
| 2 | 0.69574 | 0.26925 | 0.68033 | 0.25810 |
| 3 | 7.71486 | 0.15584 | 14.04069 | 0.08263 |
| 4 | 188.01466 | 0.02849 | 294.45477 | 0.02436 |

Table 15. Linear absorption coefficients and blackbody weights for the 5-gas/cubic WSGGM in (27) – Two oxy-fuel environments

| | 65% $CO_2$ | | 90% $CO_2$ | |
|---|---|---|---|---|
| $i$ | $k_i(1/m)$ | $a_i$ | $k_i(1/m)$ | $a_i$ |
| 0 | 0 | 0.52282 | 0 | 0.66567 |
| 1 | 0.42019 | 0.28898 | 0.40334 | 0.20536 |
| 2 | 9.63050 | 0.16303 | 13.92300 | 0.10516 |
| 3 | 242.96000 | 0.02517 | 351.06000 | 0.02381 |

Table 16. Linear absorption coefficients and blackbody weights for the 4-gas/cubic WSGGM in (28) – Two oxy-fuel environments